\PassOptionsToPackage{table,xcdraw}{xcolor}
\PassOptionsToPackage{hyphens}{url}

\documentclass[acmsmall,natbib=true,anonymous=false,screen=true]{acmart}

\usepackage[table]{xcolor}
\usepackage{type1cm}     %
\usepackage{graphicx}     %
\usepackage{xspace}     %
\usepackage{blkarray}
\usepackage{booktabs}     %
\usepackage[show]{chato-notes}
\usepackage{enumitem}
\usepackage{siunitx}          %
\usepackage{mathtools}     %
\usepackage{multirow}
\usepackage{nicefrac}
\usepackage{amsmath}     %
\usepackage[utf8]{inputenc}
\usepackage{hyperref}

\newcommand{\R}[1]{{\small \href{https://reddit.com/r/#1}{\texttt{r/#1}}}}
\newcommand{\donald}[0]{\R{The\_Donald}\xspace}

\newcommand{\rev}[1]{{{#1}}}

\newcommand{\linkmatrix}[4]{
  \raisebox{1.0\baselineskip}{
      \renewcommand{\arraystretch}{1.6}
      \begin{blockarray}{c@{}ccc}
          &&\BAmulticolumn{2}{c}{\text{\small Target}}\\[-2\jot]
          && T & C \\
          \begin{block}{c@{\hspace{.6em}}c(cc)}
              \multirow{2}{*}{\rotatebox[origin=c]{90}{\text{\small Author}}}
              & T & #1 & #2 \\
              & C & #3 & #4 \\
          \end{block}
      \end{blockarray}
  }
}

\newcommand{\leaning}[1]{\ensuremath{L_{#1}}\xspace}

\author{Gianmarco De~Francisci~Morales}
\affiliation{\institution{ISI Foundation, Italy}}
\email{gdfm@acm.org}

\author{Corrado Monti}
\affiliation{\institution{ISI Foundation, Italy}}
\email{corrado.monti@isi.it}

\author{Michele Starnini}
\affiliation{\institution{ISI Foundation, Italy}}
\email{michele.starnini@isi.it}

\usepackage{fancyhdr}
\begin{document}

\acmISBN{2045-2322}
\setcopyright{acmcopyright}
\acmDOI{10.1038/s41598-021-81531-x}
\acmConference[Sci Rep 11]{Scientific Reports}
\copyrightyear{2021}
\acmPrice{}
\acmVolume{11}
\acmNumber{2818}
\acmYear{2021}
\acmMonth{2}
\def\acmBooktitle#1{\gdef\@acmBooktitle{#1}}
\acmBooktitle{Scientific Reports}

\makeatletter
\renewcommand{\@permissionCodeOne}{2045-2322}
\makeatother

\makeatletter
\renewcommand{\@copyrightowner}{Authors}
\makeatother

\makeatletter
\renewcommand{\@copyrightpermission}{%
Published on: \emph{Scientific Reports} volume 11, Article number: 2818 (February 2021). %
DOI: \href{https://doi.org/10.1038/s41598-021-81531-x}{10.1038/s41598-021-81531-x} %
\medskip \\ %
This article is licensed under a Creative Commons Attribution 4.0 International License, which permits use, sharing, adaptation, distribution and reproduction in any medium or format, as long as you give appropriate credit to the original author(s) and the source, provide a link to the Creative Commons licence, and indicate if changes were made. The images or other third party material in this article are included in the article's Creative Commons licence, unless indicated otherwise in a credit line to the material. If material is not included in the article's Creative Commons licence and your intended use is not permitted by statutory regulation or exceeds the permitted use, you will need to obtain permission directly from the copyright holder. To view a copy of this licence, visit \href{http://creativecommons.org/licenses/by/4.0/}{http://creativecommons.org/licenses/by/4.0/}.}
\makeatother

\title{No Echo in the Chambers of Political Interactions on Reddit}

\begin{abstract}
Echo chambers in online social networks, whereby users' beliefs are reinforced by interactions with like-minded peers and insulation from others' points of view, have been decried as a cause of political polarization. 
Here, we investigate their role
in the debate around the 2016 US elections on Reddit, a fundamental platform for the success of Donald Trump. %
We identify Trump vs Clinton supporters  %
and reconstruct their political interaction network.
We observe a preference for cross-cutting political interactions between the two communities rather than within-group interactions, thus contradicting the echo chamber narrative.
Furthermore, these interactions are asymmetrical: Clinton supporters are particularly eager to answer comments by Trump supporters. %
Beside asymmetric heterophily, users show assortative behavior for activity, %
and disassortative, asymmetric behavior for popularity. %
Our findings are tested against a null model of random interactions, by using two different approaches:
a network rewiring which preserves the activity of nodes, and a logit regression which takes into account possible confounding factors.
Finally, we explore 
possible socio-demographic implications. %
Users show a tendency for geographical homophily %
and a small positive correlation between cross-interactions and voter abstention.
Our findings shed light on public opinion formation on social media, 
calling for a better understanding of the social dynamics at play in this context. %
\end{abstract}

\maketitle

\flushbottom
\thispagestyle{empty}

\section{Introduction}

Polarization is a defining feature of contemporary politics~\citep{pew2017partisan}.
Polarization along party lines in the United States is on the rise~\citep{baldassarri2008partisans}, and 2016 elections have deepen the divide~\citep{jacobson2016polarization}.
This polarization is easy to observe online, and especially on social media, where people share their opinions liberally.
Indeed, several platforms have been the subject of polarization studies, from Twitter, to YouTube, to Facebook~\citep{conover2011political,garimella2016quantifying,an2014partisan,bessi2016users}.
\rev{On Twitter, in particular, political polarization can be exacerbated by social bots that amplify divisive messages \citep{caldarelli2020role}. }
Polarized issues fall not only along ideological fault-lines~\cite{conover2011political}, but can also touch on any collectively resonant topic~\citep{garimella2018quantifying}.

Several scholars have identified social media itself as a cause of polarization, citing ``echo chambers'' as a cause~\citep{garrett2009echo,gilbert2009blogs,quattrociocchi2016echo}.
Echo chambers are situations in which users have their beliefs reinforced due to repeated interactions with like-minded peers and insulation from others' points of view~\citep{garimella2018political,cinelli2020echo}.
The dynamics leading to echo chambers on online social networks have been associated to selective exposure~\citep{klapper1960effects}, biased assimilation~\citep{lord1979biased}, and group polarization~\citep{baumann19}; \rev{in particular, \citet{garrett2009echo} pointed to the pursuit of opinion reinforcement as a possible cause.}
Echo chambers have been empirically observed and characterized around several controversial topics, such as abortion or vaccines~\citep{garimella2018political, cossard2020falling}.
Many have expressed concern that, as citizens become more polarized about political issues, they do not hear the arguments of the opposite side, but are rather surrounded by people and news sources who express only opinions they agree with (e.g., Mark Zuckerberg~\cite{zuck}).
However, the very existence of such echo chambers has been recently questioned~\citep{dubois2018echo,guess2018avoiding}, as well as their relation with the news feed algorithm of different social media platforms \cite{cinelli2020echo}.
In particular, the effect of echo chambers in increasing political polarization has been put under scrutiny~\citep{guess2018avoiding,bail2018exposure}.

In this paper, we consider a highly polarized issue, the 2016 US presidential elections, and investigate the role of echo chambers on social media in exacerbating the debate. 
We focus on Reddit as a platform where to study political interactions between groups with opposite views.
Reddit is a social news aggregation website, in 2016, it was the seventh most visited website in United States, with more than 200 million visitors.
It was a fundamental platform for the success of Donald Trump's political campaign~\cite{karpf2017digital}.
Given this context, we set out to {characterize the interaction patterns between opposing political communities on Reddit}, by considering supporters of the two main presidential candidates, Clinton and Trump.
Then, we look at the way they interact in a common arena of political discussion (i.e., the most popular subreddit related to politics), by reconstructing the information flow between users, as determined by their comments and replies. 
Are echo chambers responsible for the increased polarization on Reddit during the 2016 electoral cycle~\citep{nithyanand2017online}?

Our empirical investigation shows that %
there is {no evidence of echo chambers} in this case. 
On the contrary, cross-cutting political interactions between the two communities are more frequent than expected. 
This heterophily is not symmetric with respect to the two groups: Clinton supporters are particularly eager to answer comments by Trump supporters, an asymmetry that is not explained by other confounders.
Finally, we ask how these findings are modulated by socio-demographics, environmental characteristics of the Reddit users involved in the discussions, determined by geo-localization of such users.
Our results point at a preference for geographical homophily in online interactions: users are more likely to interact with other users from their own state.
We find a statistically significant (albeit small) positive correlation between cross-interaction and voter abstention, which may support the hypothesis that exposure to cross-cutting political opinions is associated with diminished political participation~\cite{bakshy2015exposure,mutz2002consequences,huckfeldt2004disagreement}.
We obtain a similar result, although in the negative direction, for living in a swing state: cross-interactions are suppressed in this case.

These results have important implications in terms of our understanding of public opinion formation.
It is often assumed that echo chambers can be pierced by increasing the amount of cross-cutting content and interactions between the polarized sides~\citep{bakshy2015exposure,garimella2017reducinga}.
\rev{Instead, the present study shows that polarization around a highly controversial issue, such as 2016 U.S. Presidential elections~\cite{johnston2018increasingly}, can co-exist with a large presence of cross-cutting interactions.}
The nature of these interactions might even increase polarization via ``backfire effect''~\citep{nyhan2010corrections}, as recently empirically found for Twitter~\citep{bail2018exposure}.
Alternatively, cross-cutting interactions and polarization might be the result of growing underlying socio-economical divisions~\cite{duca2016income,storper2018separate}.
Overall, our findings call for a better understanding of the social dynamics at play in this context before suggesting technical solutions for such social systems, which could have unintended consequences.

\section{Political interactions on Reddit}
\label{sec:data}

\rev{
We gather data from Reddit, through the Pushshift collection~\cite{baumgartner2020pushshift}.
Reddit is organized in communities, called \emph{subreddits}, that share a common topic and a specific set of rules.
Users subscribe to subreddits, which contribute to the news feed of the user (their \emph{home}) with new posts.
Inside each subreddit, a user can \emph{post}, or \emph{comment} on other posts and comments.
Thus, the overall discussion under each post evolves as a tree structure,  growing over time.
In addition, users can also \emph{upvote} posts and comments to show approval; they manifest disapproval with a \emph{downvote}.
Each message therefore is associated to a \emph{score}, which is the number of upvotes minus the number of downvotes it has received. %
}

Given the two-party nature of the US political system and the polarized state of its political discourse, we approach the problem by modeling the interactions between groups of users labeled by their political leaning---specifically, according to which candidate they support in the 2016 presidential elections.
We then model such interactions as a weighted, directed network, where nodes represent users and links represent comments between them.
On top of political leaning, we also characterize the users in terms of their activity, i.e., their propensity to engage in interactions with other peers, and popularity, as given by the score assigned to their comments.
In the remainder of this section we explain these three steps more in detail.

\subsection*{Political leaning of Reddit users}
\label{sec:leaning}

We identify the political leaning of Reddit users by looking at their posting behavior.
With respect to the 2016 US presidential elections,  users can be characterized as supporting the Democratic candidate, Hillary Clinton, or the Republican candidate, Donald Trump. 
On Reddit, we identify specific subreddits dedicated to supporting the main presidential candidates.
For Donald Trump, we select the subreddit \donald; for Hillary Clinton, we choose the subreddits \R{hillaryclinton} and \R{HillaryForAmerica}.

The subreddit \donald was created in June 2015, at the beginning of Donald Trump campaign for the Republican party nomination.
It has been one of the largest online communities of Trump supporters, with \num{269904} users in November 2016. %
Participation in this subreddit is a valid proxy to study Donald Trump support, as the rules of this subreddit explicitly state that the community is for ``Trump Supporters Only'', and that dissenting users will be removed.
As such, it has been previously used in literature to analyze the behavior of Trump supporters~\cite{flores2018mobilizing,rootsoftrumpism}.
\R{hillaryclinton} and \R{HillaryForAmerica} are the main communities that supported the Hillary Clinton's campaign in 2016.
The former was created in 2015, while the latter was created in 2016 specifically to support her presidential bid.
In November 2016, they were able to attract \num{35142} and \num{3025} Reddit users, respectively.
Since the stated goal of these communities is to support her presidential campaign, and they forbid the use of the subreddit to campaign for other candidates, we consider active participation in these communities as a good proxy for support for the Democratic party candidate.
We call these subreddits the \emph{home communities} for each candidate.
We identify \num{117011} users who actively posted on \donald in 2016, and \num{13821} on \R{hillaryclinton} and \R{HillaryForAmerica}.
Given the massive use of Reddit as a political tool by Trump's campaign~\cite{karpf2017digital}, the difference in size between the two communities is not surprising.

Although these subreddits are dedicated to supporters of the candidate, we find that \num{3702} users post in both subreddits ($2.9\%$).
In order to disambiguate the leaning for these users, we retrieve the Reddit score of their comments in the home communities.
The score represents the difference between the number of upvotes and the number of downvotes assigned by other users visiting the same subreddit.
Upvotes are generally understood to encode approval, appreciation, or agreement; downvotes encode their opposites.
Thus, a user with a higher score on Clinton and a lower score on Trump is most likely a Democratic supporter. 
Following this reasoning, among all users who posted on both home communities, we consider a user as Clinton supporter if they have an average score on their comments on the Clinton home community that is larger than the score on the Trump home community and vice-versa.
Users with tied scores are discarded, as they represent only $5\%$ of the set of tied users ($0.145\%$ of the overall set of users).

Therefore, we define the political leaning of a user $u$ as a binary label \leaning{u}, assigned as Clinton supporter ($\leaning{u}=C$), if they post only on Clinton's home community, or they posts on both communities and have a larger average score on Clinton's community, and as Trump supporters otherwise ($\leaning{u}=T$).
Our method identifies \num{10240} users as Clinton supporters %
and \num{110806} users as Trump supporters.

\subsection*{Network of interactions on Politics}
\label{sec:net}

To study the interactions between the two sides, we need a community that is visited regularly by both groups, but which is still topically related to politics and popular enough.
The best candidate for such a role is \R{politics}, since it is the largest political subreddit.
We collect all submissions and comments in the year 2016.
From the collected comments, we reconstruct the network of political interactions among the users we previously identified.
Among these users, \num{31218} authored a message on \R{politics} in 2016 and thus appear as nodes $V$ in the graph ($N_T=\num{27012}$ Trump supporters, $N_C=\num{4206}$ Clinton supporters).
Nodes correspond to users with known political leaning, while a weighted, directed link $(u,v)$ corresponds to user $u$ posting a comment as a response to user $v$.
The weight $w_{uv}$ corresponds to the number of such interactions from $u$ to $v$.
Note that the link direction represents the interaction, and is opposite to the information flow (user $u$ should have read what $v$ wrote to answer, but it is not guaranteed that $v$ will read $u$'s reply).

\begin{table}[tb]
  \centering
  \caption{Main properties of the Politics network: number of users $N$, divided in Trump/Clinton supporters $N_T$ / $N_C$, number of links $E$, average degree $\langle k \rangle$, reciprocity $\rho$ (fraction of bidirectional links over the total), and total number of interactions  $W$.  }
  \label{table:graph-properties}
  \begin{tabular}{ccccccc}
    \toprule
    $N$ &  $N_T$ & $N_C$ & $E$ & $\langle k \rangle$ & $\rho$ & $W$    \\
    \midrule
    \num{31218}   & \num{27012} & \num{4206} & \num{500030} & \num{16.02} & \num{0.4482} & \num{716765}	 	  \\
    \bottomrule
  \end{tabular}
\end{table}

In the Politics network, the probability to find a node labelled as $X \in \{C,T\}$ (henceforth, $X$ node for brevity) in the network is $P(X) = N_X/N$, corresponding to $P(T) \simeq 0.87$ for Trump, $P(C) \simeq 0.13$ for Clinton.
The main properties of the Politics network are reported in Table~\ref{table:graph-properties}.
The joint probability to observe an interaction from an $X$ node to a $Y$ node reads
\begin{equation}
\label{eq:emp_links}
P(X \rightarrow Y) = \frac{W_{XY}}{W}
     \simeq \qquad 
	\linkmatrix{0.40}{0.25}{0.25}{0.10},
 \end{equation} 
where the rows of the matrix indicate the leaning of the author of a comment, and the columns the one of the target, $W = \num{716765}$ is the total weight of the links in the network (that is, the number of interactions between all considered nodes), and $W_{XY}$ is the weight of directed links from $X$ nodes to $Y$ nodes:
\begin{equation*}
	W_{XY} = \sum\limits_{u,v \in V \mid \leaning{u}=X \wedge \leaning{v}=Y} w_{uv}.
\end{equation*}
We denote with $W_{\rightarrow X} = \sum_Y W_{YX}$ the number of interactions received by $X$ nodes ($\sum_Y$ denotes the sum over all possible label assignments to Y), and $W_{X \rightarrow}$ the ones originated by $X$ nodes.
It follows that $\sum_{XY} W_{YX} = \sum_{X}W_{X \rightarrow} = \sum_{X}W_{\rightarrow X} = W$.

Diagonal elements of the matrix in Eq. \eqref{eq:emp_links} correspond to the interactions within political groups, off-diagonal to those across groups. 
The sum by rows (columns) of the matrix in Eq.~\eqref{eq:emp_links} corresponds to the probability that an $X$ node initiates (receives) an interaction, $P(X \rightarrow) = \frac{W_{X \rightarrow}}{W}\,$ ($P(\rightarrow X ) = \frac{W_{\rightarrow X}}{W}$).
From Eq.~\eqref{eq:emp_links}, interactions across communities, or \emph{cross-interactions} looks symmetric between Clinton and Trump communities.
However, joint probabilities do not take into account the difference in size between the two groups.
This result stems from the fact that the probability that Clinton nodes initiate an interaction, $P(C \rightarrow) = W_{C \rightarrow}/W \simeq 0.35$ is much larger than the fraction of Clinton supporters in the network, $N_C/N \simeq 0.13$, which implies that Clinton supporters have much larger weighted out-degree than Trump ones.

These characteristics can be further inspected by considering the conditional probability to observe an interaction from an $X$ node to a $Y$ node, given that the first node has leaning $X$,
\begin{equation}
\label{eq:cond_prob}
P(X\rightarrow Y \vert X)
 = \frac{P(X\rightarrow Y)}{P(X \rightarrow)}
  = \frac{W_{XY}}{W_{X \rightarrow}}
 \quad \simeq 
    \linkmatrix{0.62}{0.38}{0.72}{0.28}.
\end{equation} 

By looking at the columns of Eq.~\eqref{eq:cond_prob}, in absence of homophilic or heterophilic effects, one would expect elements of each column to be equal: given the author of a comment, the probability to interact with the two groups would be equal, given only by the size of the group.
Instead, we can observe that Clinton supporters tend to interact more with Trump supporters ($72\%$ of interactions) than Trump supporters themselves within the community ($62\%$).
The same effect is visible for Trump supporters, who are more likely to interact with Clinton ones ($38\%$ of interactions) than the Clinton community within itself ($28\%$ of interactions).
These intuitions will be solidified in Section \ref{sec:null-model}, by comparing these values to a null model of random social interactions.  

Finally, we compare the average sentiment polarity of each type of interaction.
To do so, first we measure the sentiment polarity (ranging from $-1$ to $1$) of the textual content of each interaction according to \textsc{VADER}~\cite{hutto2014vader}; then, we compute the average values according to the possible pairs of labels.
In this way, we obtain:
\begin{equation}
  \linkmatrix{1.26}{0.72}{1.10}{5.75} \times 10^{-2}.
\end{equation}
First, we observe 
that interactions within Trump supporters are more negative than interactions within Clinton supporters (average sentiment of $0.0575$ vs $0.0126$).
In addition, cross-cutting interactions between groups have on average a more negative sentiment than interactions within groups.
That is, Clinton supporters commenting Trump supporters have an average sentiment of $0.0110$, while when commenting on other Clinton supporters the average sentiment is $0.0575$.
The same is true for Trump supporters.
This difference is consistent with the hypothesis that cross-cutting interactions are a potential expression of conflict.

\subsection*{Reddit score and activity of users}
\label{sec:activity}
 
Political interactions on Reddit can be further characterized in terms of the score assigned to each comment or submission, and the activity of users, i.e., their propensity to engage in interactions with other peers.

\begin{figure}[tbp]%
    \centering
    \includegraphics[width=0.45\linewidth]{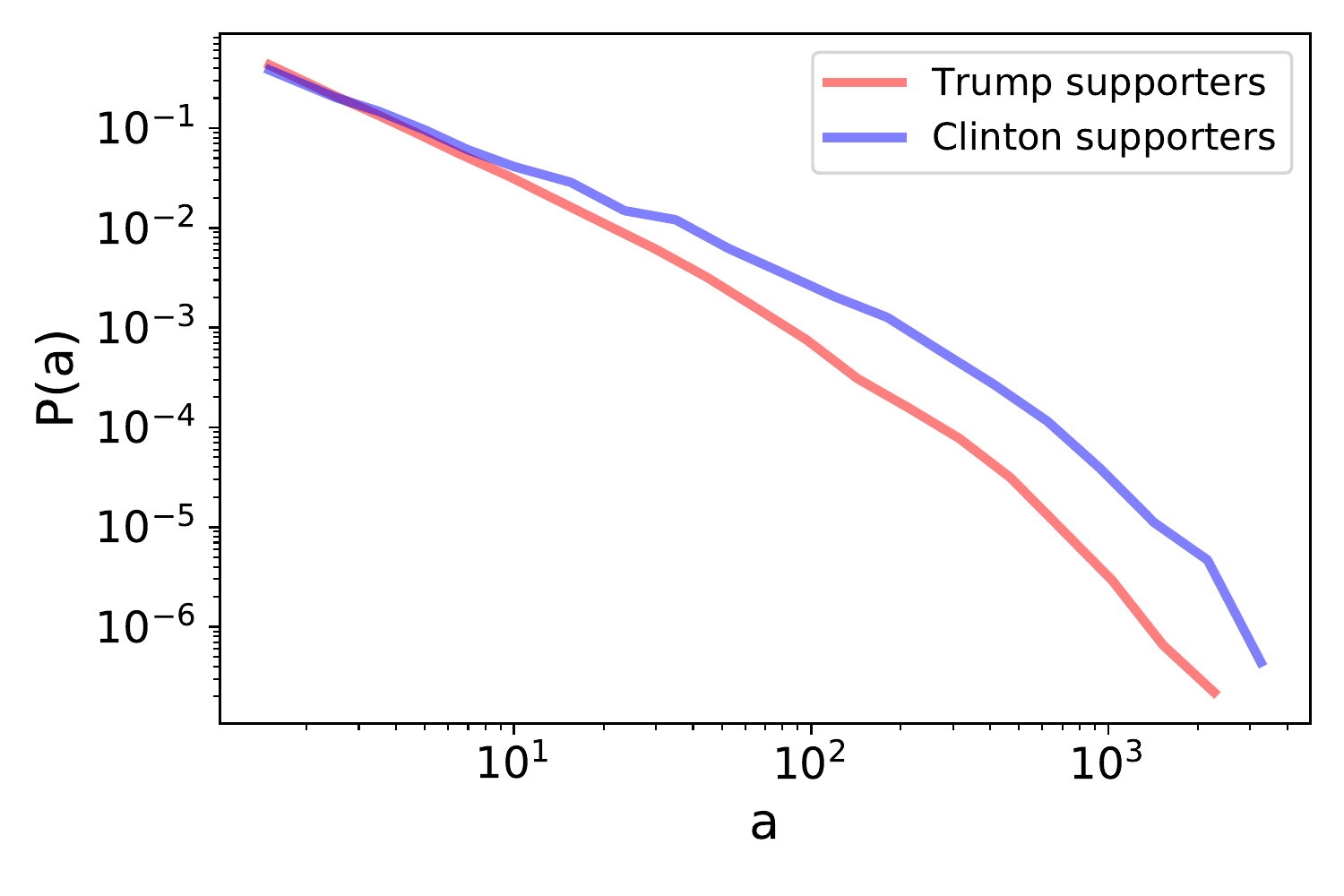}
    \includegraphics[width=0.45\linewidth]{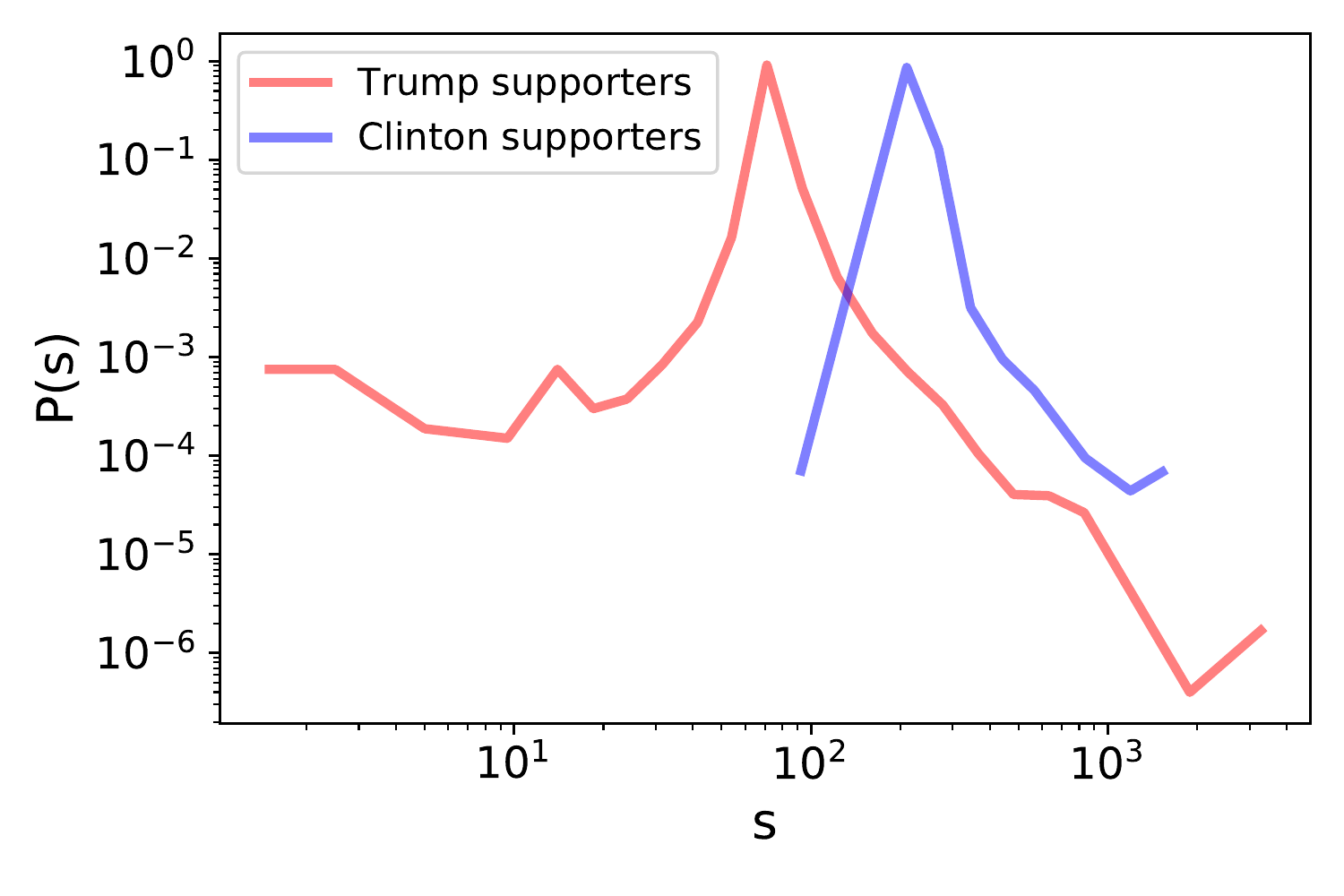}%
    \caption{(a) Probability distribution of the activity $a$ of users in the Politics network, $P(a)$, plotted separately for Clinton and Trump supporters. (b) Probability distribution of the average score $s$ of users in the Politics network, $P(s)$, plotted separately for Clinton and Trump supporters.}%
    \label{fig:activity_score}%
\end{figure}

In network terms, the activity of a user $u$, $a_u$, can be measured by the total weight of out-going links from node $u$, which corresponds to the out-strength of node 
$u$: $a_u =\sum_v w_{uv}$
Figure~\ref{fig:activity_score} (a) shows the activity distribution $P(a)$ in the Politics network, plotted separately for Clinton and Trump supporters, both with typical heavy-tailed behavior. %
The activity distribution of Trump supporters decays more rapidly than for the Clinton ones, thus indicating a propensity to engage in a larger number of interactions from Clinton supporters.
 
The Reddit score of a comment is a measure of its popularity and, as such, it  strongly depends on the subreddit where this comment is posted: popular comments posted on the subreddit \donald will be likely unpopular in subreddit where opposite political views dominate, such as Clinton-oriented subreddits. 
We define the popularity of a user $u$ on a subreddit as the average score of their comments on that subreddit, $s_u$, and it will thus depend on the subreddit under consideration. 
Figure~\ref{fig:activity_score} (b) shows the popularity distribution $P(s)$ of users in the Politics network, separately for Clinton and Trump supporters.
While the function form of the $P(s)$ distribution is similar for Clinton and Trump supporters, comments by Clinton supporters have much larger scores on average, while the scores of Trump supporters span a larger interval of values.
This observation implies that the overall attitude on the politics subreddit is more favorable to comments from Clinton than from Trump supporters, although users classified as Trump supporters are a much larger set than Clinton supporters.

This liberal bias in the general opinion of \R{politics}, however, does not seem to discourage Trump supporters from commenting in large numbers.
Therefore, since we wish to study the two communities and how they interact, \R{politics} is the best arena to observe such interactions.
Our set of users of interest is not a representative of \R{politics} users.
Nevertheless, we are not interested in studying the typical behavior of users in this subreddit, but in analyzing how these two polarized communities interact in this arena.
The fact that the two communities are not representative of the politics subreddit is therefore of no consequence.

\section{Comparison with a null model of random interactions}
\label{sec:null-model}

To understand whether the empirical patterns observed in the previous section represent a consistent behavior, we need to compare them with a theoretical null model of interactions.
The simplest null model for our data follows the hypothesis that the interactions are unaffected by the political leaning of users.
In mathematical terms, the null model is a directed, weighted, random network (RN). 
This network is obtained by reshuffling links of the original network while preserving the in- and out- strength (weighted degree) of each node.

In this network, the probability to observe a link from an $X$ node to a $Y$ node is the product of two independent probabilities: the probability that an $X$ node initiates an interaction, and the probability that a $Y$ node receives an interaction,
\begin{equation}
\label{eq:null}
p_{RN}(X \rightarrow Y) = 
\frac{W_{X \rightarrow}W_{\rightarrow Y}}{W^2}
\simeq \quad \linkmatrix{0.43}{0.225}{0.225}{0.12}.
\end{equation} 
The RN model preserves both the in- and out-strength sequence of nodes, while rewiring connections among them, thus following the so called \emph{configuration model}~\citep{molloy1995critical}.
In this RN model, the conditional probability to observe a link from an $X$ node to a $Y$ node, given that the first node has leaning $X$, reads
\begin{equation}
\label{eq:cond_null}
p_{RN}(X\rightarrow Y \vert X) =
\frac{W_{\rightarrow Y}}{W}
\simeq \quad
    \linkmatrix{0.34}{0.66}{0.34}{0.66}.
\end{equation}

In the following, we investigate deviations of observed data from this RN model, so to highlight specific patterns of behavior.

\begin{figure}[tbp]%
    \centering
    \includegraphics[width=0.45\linewidth]{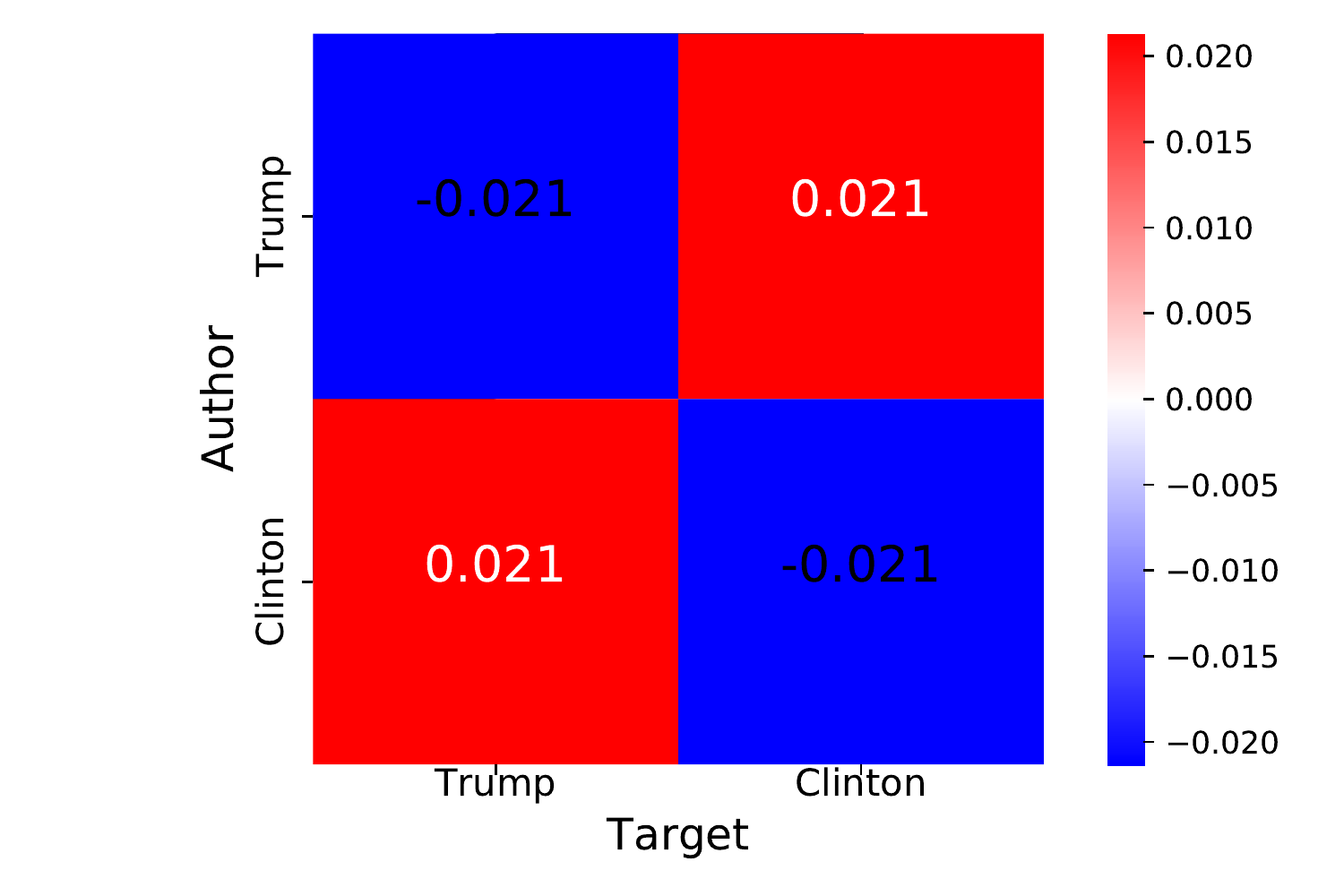}
    \includegraphics[width=0.45\linewidth]{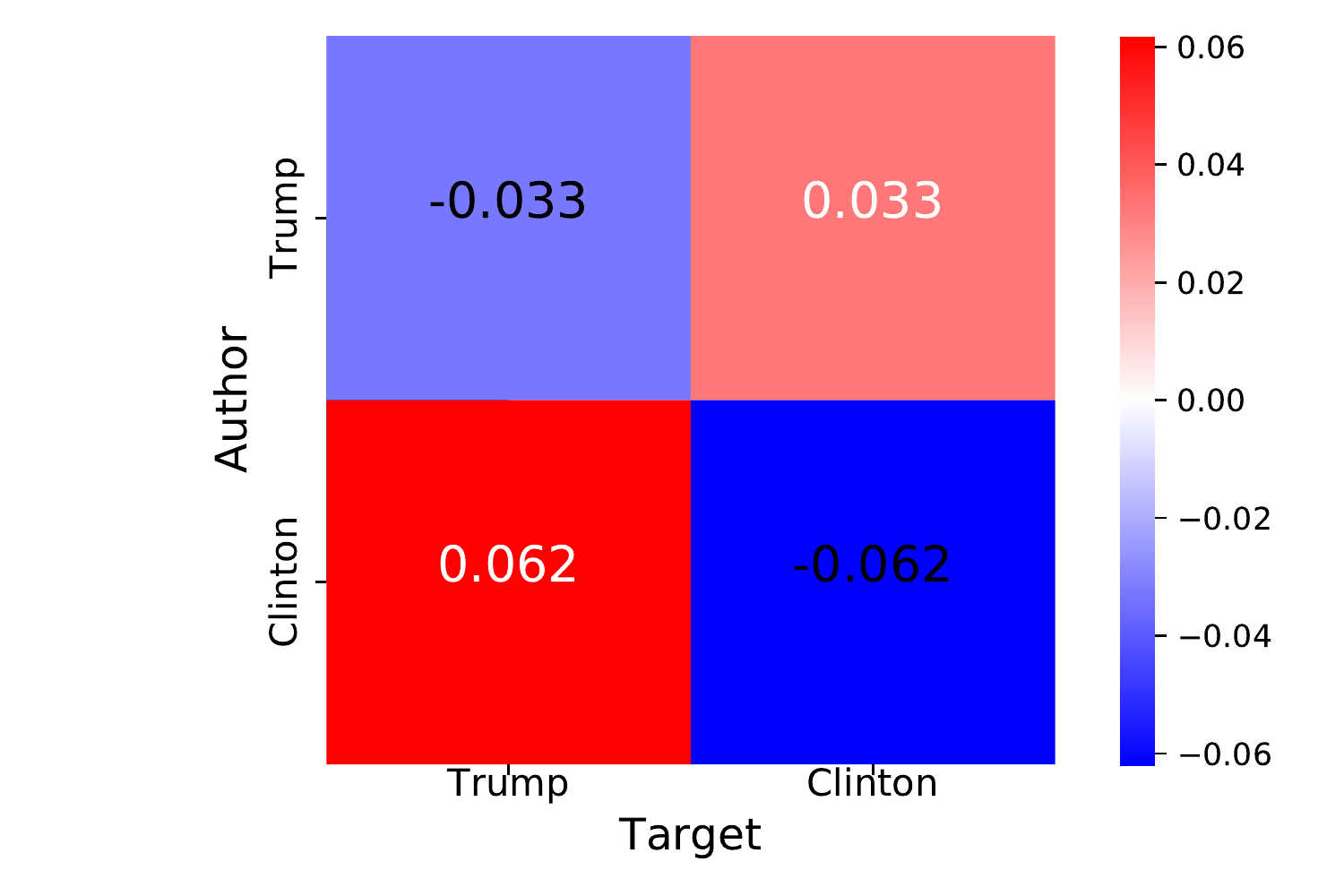}%
    \caption{\rev{
    Difference between empirical and random joint probabilities (left) and between empirical and random conditional probabilities (right) of interaction in the Politics network, with respect to Trump and Clinton supporters. 
    We observe heterophily and asymmetry: off-diagonals are larger than zero and show distinct values (right-side plot).}
    }%
    \label{fig:diff_leaning}%
\end{figure}

\subsection*{Deviation of conditional and joint probabilities}

The difference between the empirical and random joint probabilities, given by Equations~\eqref{eq:emp_links} and~\eqref{eq:null}, respectively, is shown in Figure~\ref{fig:diff_leaning}~(a). 
Cross-interactions between opposite political groups in the Politics networks happen more often than expected in a RN model\rev{, with an odds ratio of 1.195}.
This observation implies that there is a certain degree of heterophily in interactions, i.e., the preference to interact with users from the the opposite political group.
This result is surprising, considering the ample literature about homophily in social networks and especially about echo chambers in political discussion on social media~\citep{conover2011political,garimella2018political}.
The difference between empirical and random conditional probabilities, given by Equations~\eqref{eq:cond_prob} and~\eqref{eq:cond_null}, respectively, is reported in Figure~\ref{fig:diff_leaning} (b).
The Politics network is characterized by an asymmetry between the two political groups: Clinton supporters interact with Trump supporters more than the other way around, \rev{with a $6.2\%$ increase with respect to a random null model on one side and $3.3\%$ on the other. }

Given that Trump and Clinton supporters are different according to several metrics, there may be some confounding effects in the interactions.
In particular, we explore the roles of activity (the strength of the node) and popularity (the average score of the node).
To give a visualization similar to the ones in Figure~\ref{fig:diff_leaning}, we define activity and score classes from the distributions shown in Figure~\ref{fig:activity_score}. 
We manually define 4 score classes, from low to high score, and 8 activity classes, from low to high activity. 
In the following we indicate for brevity a user of score class $s$ as a $s-$user, and a user of activity class $a$ as a $a-$user.  

\begin{figure}[tbp]%
    \centering
    \includegraphics[width=0.45\linewidth]{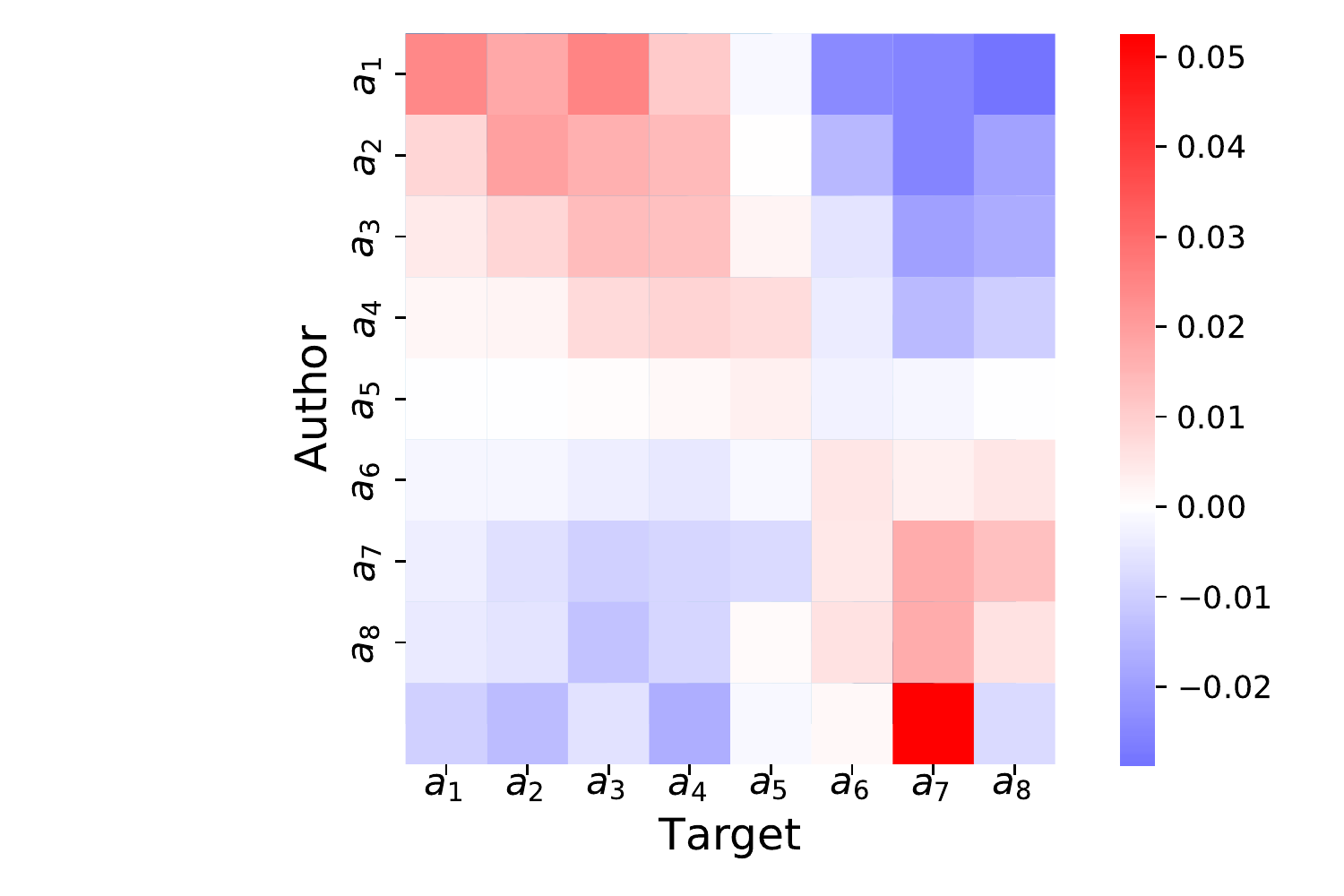}
    \includegraphics[width=0.45\linewidth]{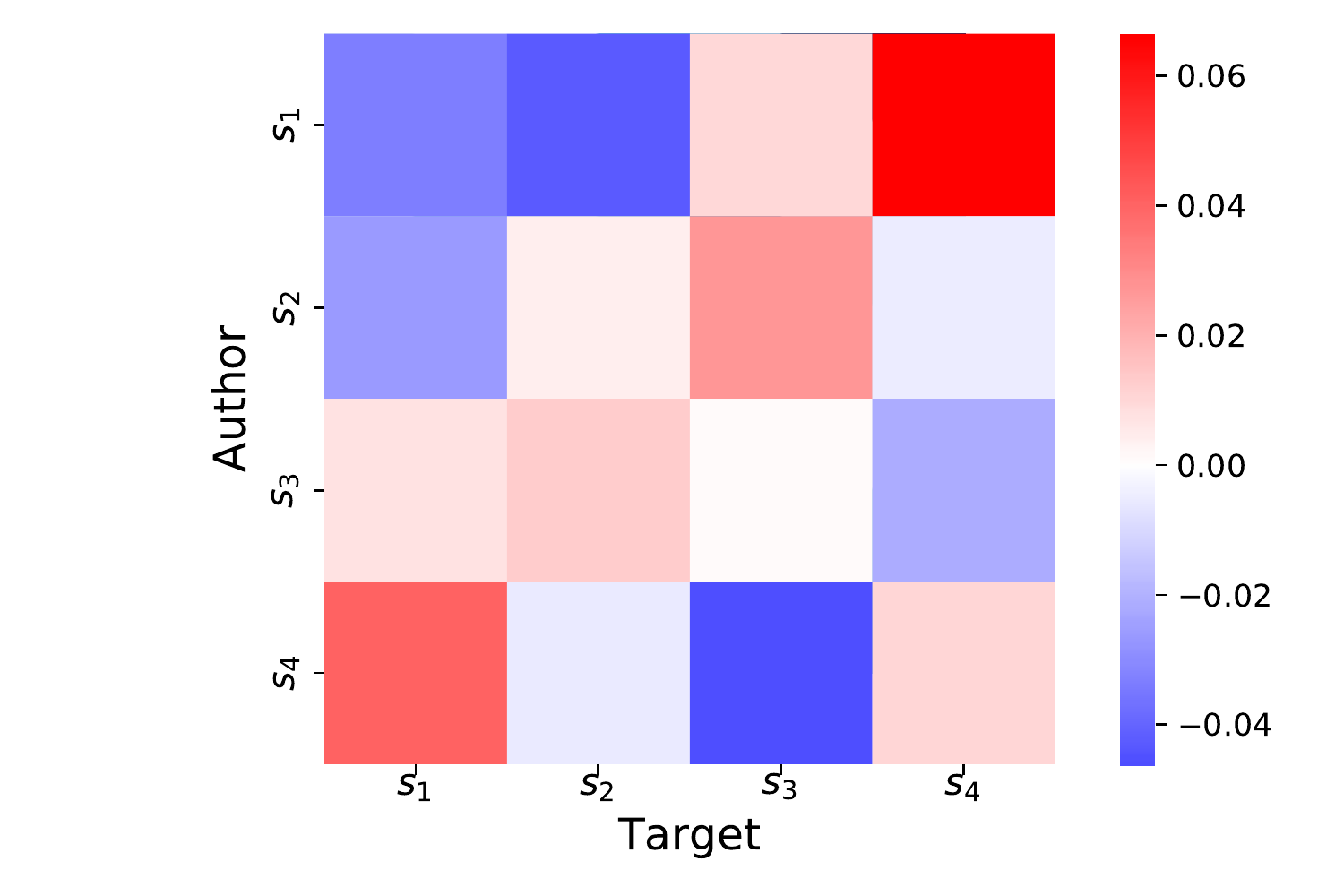}%
    \caption{
    \rev{Difference between empirical and random conditional probabilities in the Politics network with respect to activity classes (on the left, with $a_1$ being the least active users) and to score classes (on the right, with $s_1$ being the lowest scores). %
    We observe assortative behavior with respect to activity (higher values on diagonal) and disassortative behavior with respect to score (higher values off diagonal).}
    }%
    \label{fig:diff_activity_score}%
\end{figure}

Next, we define the empirical  probability  to observe a link from an $a-$node to an $a'-$node,
$P(a\rightarrow a')$, by applying Eq. \eqref{eq:emp_links} to activity classes, i.e.,
$P(a\rightarrow a')= W_{a,a'}/W$ where $W_{a,a'}$ is the number of interactions from $a-$nodes to $a'-$nodes. 
The same can be done for the conditional probability, by applying Eq. \eqref{eq:cond_prob} to activity classes. 
For random interactions represented by the null model, one can obtain the joint probability $P_{RN}(a\rightarrow a')$ and conditional 
probability $P_{RN}(a \rightarrow a' \vert a)$.  
Figure~\ref{fig:diff_activity_score} (left) shows the difference between empirical and random conditional probabilities for interactions with respect to activity classes. \rev{Positive values indicate that a pair of classes interacts more than expected by random chance; negative values indicate that they interact less than expected.}
Thus, we observe that users show an assortative behavior with respect to activity, i.e., user with high activity tend to interact with similarly-active users, and the same for users with low activity.

We can also define the probability to observe a link from a $s-$node to a $s'-$node,
$P(s\rightarrow s')$, by applying Eq. \eqref{eq:emp_links} to score classes.
The same can be done for the conditional probability and for the null model. 
Figure~\ref{fig:diff_activity_score} (right) shows the difference between empirical and random conditional probabilities for interactions with respect to score classes. 
Users show a disassortative, slightly asymmetric behavior with respect to scores: users with low score tend to interact with users with high score, viceversa is less slightly frequent. 
This is due to popular comments attracting many comments from other users, mostly of low popularity. %

 Given these characteristics of the network, it is important to understand how the differences in behavior w.r.t. activity and score affect the heterophily and asymmetry results found for the leaning.
 To do so, we need a unified model that puts all these ingredients together, and confronts the resulting model vs the random network null model.
 The next section explains our approach to tackle this task.
 
 \subsection*{Logit regression model}
 \label{sec:logit}

So far, we have recognized the effects of different groups and user characteristics in the commenting behavior.
We now wish to assess whether the effects we have identified so far are statistically significant, and what is the relationship among them.
In particular, we want to see how the variables of interest for our study (community interactions) are confounded by the popularity variables.
There is no need to control for activity variables as our null model already takes that aspect into account, as explained next.
To do so, we design a logit model, in order to quantify the odds of within-group and cross-group interaction, and the role of each variable.
Such a model also allows us to study the effect of geographical-based variables in the next section.

Our logit model defines the probability of $u$ interacting with $v$ as a function of the features of $u$ and $v$.
We consider three sets of features: \emph{community} interaction features (the political labels we defined), \emph{confounding} Reddit features (popularity metrics), and \emph{environmental} features which capture real-world phenomena. 

For community features, similarly to the previous section, we consider the following set of binary features: 
\begin{itemize}
  \item \emph{Clinton support}: $1$ if $L_u=C$ ($u$ is a Clinton supporter), $0$ otherwise.
  \item \emph{Cross-group}: $1$ if $u$ and $v$ support different candidates ($L_u \neq L_v$), $0$ otherwise.
  \item \emph{Clinton support, Cross-group}: interaction feature between the previous variables, $1$ if $L_u=C$ and $L_v=T$, $0$ otherwise.
\end{itemize}
The combinations of these variables, that we assume to be independent, represent all four possible scenarios of political labels (depicted in Figure~\ref{fig:diff_leaning}(a)).

We then add several variables to control for potential effects of different levels of popularity between supporters of Clinton and Trump.
We operationalize these confounding features as follows:
\begin{itemize}
  \item \emph{Average score}: average score obtained, separately, by $v$ (\emph{target}) and by $u$ (\emph{author}) ($s_v$ and $s_u$).
  \item \emph{Difference in average score}: the absolute difference between the average scores of $v$ and $u$, namely $\vert s_v - s_u \vert$.
  \item \emph{Fraction of positives}: the fraction of comments with a positive score over the total number of comments, separately for $v$ (\emph{target}) and for $u$ (\emph{author}).
  \item \emph{Difference in fraction of positives}: the absolute difference between the fractions of $u$ and $v$.
\end{itemize}
We quantile-normalize these features so that they all display the same distribution, thus allowing to easily interpret the coefficients of our logit regression model.

We choose to use both the average score and the fraction of comments with positive score because the scores have a heavy tailed distribution, therefore the average might be skewed.
Conversely, the fraction of positives represents a summary statistic on a boolean property, and thus captures a different aspect of the data.
We also add the differences to include link-based features that capture the dynamics of the interaction between the specific author and target.

To model the effects of these variables on comment creation, we use a logistic regression model.
In this data set, we have $N=\num{31218}$ users and $W=\num{716765}$ comments.
It is thus unfeasible to use the complete set of negative links (i.e., pairs of users $(u, v)$ where $u$ did not interact with $v$).
We therefore resort to sampling the set of negatives.
This sampling procedure only changes the value of the intercept coefficient, which is not of interest, and does not affect the interpretation of the important parts of the model (the coefficients of the community variables under study).

It is important to carefully choose the sampling strategy, as it needs to faithfully represent the null model we are considering.
In the null model we presented in Section~\ref{sec:null-model}, we consider the author and the target of the comments to be fixed, that is, we rewire the network while preserving the in- and out- strength of each node, equivalent to a configuration model~\citep{molloy1995critical}.
In the sampling strategy of negative links, we follow exactly the same procedure.
We choose node $u$ with probability proportional to their out-strength or activity $a_u$, node $v$ with probability proportional to their in-strength, defined in Section \ref{sec:activity}.
If the link between $u$ and $v$ exists, we discard it.
This way, the negative sample reflects exactly the null model presented in Section~\ref{sec:null-model}: the probability of considering a pair of nodes is just the product of two independent probabilities -- the probability that a node $u$ initiates an interaction, and the probability that a node $v$ receives it.
The role of logistic regression is thus to capture how the variables we consider alter the chances of observing a link $u \rightarrow v$.

\subsection*{Logit regression results}

First, we present results for the model that only considers the community variable.
We report the odds ratios obtained for this model in the first column of Table~\ref{table:logit-table-reddit}.
All the coefficients are statistically significant at the $0.1\%$ level.
These results confirm our analysis so far:
\begin{enumerate}[label=(\roman*)]
  \item comments on \R{politics} are heterophilic: the likelihood of $u$ answering to $v$ increases when $u$ and $v$ support different candidates (odds ratio $1.195$);
  \item Clinton supporters are less likely to leave a comment than Trump's (odds ratio $0.942$), in general;
  \item however, Clinton supporters are asymmetrically slightly more likely to leave a comment when the user they are responding to supports the other candidate (odds ratio $1.064$).
\end{enumerate}

\begin{table}[t]
  \centering
  \caption{\label{table:logit-table-reddit} %
  Odds ratios obtained by logistic regression. Each column corresponds to a model with a specific set of variables: we have \emph{community} features in the first column; then, we add to the model the first, the second, and both sets of \emph{confounding} variables.
  All the variables shown here have a statistically significant ($p < 0.001$) impact on the likelihood of writing a comment.}
  \small
  \rowcolors{2}{gray!20}{white}
  \begin{tabular}{lllll}

\toprule
Variable name &  Comm. & Conf. 1 & Conf. 2 & Conf. 1+2 \\
\midrule
    
Clinton sup.              &  0.942*** &  0.918*** &  0.936*** &  0.911*** \\
Cross-group               &  1.195*** &  1.172*** &  1.191*** &  1.165*** \\
Clinton sup., Cross-group &  1.064*** &  1.091*** &  1.070*** &  1.102*** \\
Avg. score (author)       &           &  1.166*** &           &  1.174*** \\
Avg. score (target)       &           &  1.151*** &           &  1.167*** \\
Diff. avg. score          &           &  1.213*** &           &  1.228*** \\
Diff. frac. positive      &           &           &  0.497*** &  0.498*** \\
Frac. positive (author)   &           &           &  1.260*** &  1.247*** \\
Frac. positive (target)   &           &           &  1.221*** &  1.195***
\\ %
\bottomrule
\multicolumn{5}{c}{* $p < 0.05$, *** $p < 0.001$}
\end{tabular}

\end{table}

We can also compute the interaction matrix obtained from this model.
To do so, we multiply the odds ratios obtained by the model for each of the four possible combinations of groups between author and target.
In this way, we obtain
\begin{equation}
  \linkmatrix{0.925}{1.105}{1.107}{0.871}
\end{equation}
The results obtained via this methodology are in line with those presented in the previous section (i.e., Figure~\ref{fig:diff_leaning}).

Then, we control for the other Reddit variables we analyzed, in order to assess whether these effects are robust or if they can be explained by considering these other features.
We report results for models that include the average score, the fraction of positively scored comments, and both, in the other columns of Table~\ref{table:logit-table-reddit}.
Including these variables do not affect the neither the odds ratios nor the statistical significance of the community features.
This result confirms that the effects we observe for political interactions across communities are not confounded by these other user characteristics.

These control variables, in addition, show that users with a higher average score are more likely both to initiate and receive interactions.
This relationship is heterophilic: a large difference in average score \rev{is associated to} an increased likelihood.
We can interpret average score as a proxy measure for visibility: authors able to attract a large number of upvotes are also more prolific, and they also attract more comments, which explains the larger-than-one ratios.
They also tend to trigger a response even from unpopular authors, thus explaining the heterophily.
Again, the results obtained with the logit regression model are in agreement with what observed by comparing empirical interactions with a random network (i.e., Figure~\ref{fig:diff_activity_score}).

Authors with a larger fraction of positively scored comments also are more likely to send and receive comments.
In fact, since score is a measure of the social feedback from the community, this shows that authors more aligned with the community tend to be more active in it, which is not surprising.
With this variable, however, the relationship is homophilic: users positively scored and users negatively scored are more likely to comment each other.
This result might be an effect of the community trying not to ``feed the trolls''~\cite{bergstrom2011don}.

\section{Sociodemographic implications} 

In this section, we investigate the connections between online interactions present in our data and offline socio-demographic factors.
In particular, our research question is the following: \emph{which environmental factors are associated to higher levels of online cross-group interactions?}
While we cannot prove any causal effect of the environmental factors, such observational study can provide insights for theory generation and followup investigation.

To answer our research question, we need a proxy of the socio-demographic environment of users. 
We choose US states as a proxy, as this is the finest spatial granularity we can reliably infer for Reddit users. 
We infer the state of each user according to the information gathered by \citet{balsamo2019firsthand}, which is based on the usage of local Reddit communities.
Out of our set of \num{121046} Reddit users, we are able to geo-localize $37\%$ of them at the state level.
Henceforth, we restrict our analysis only to comments authored by users in this set.
State information is slightly unbalanced ($36\%$ for Trump's supporters, $43\%$ for Clinton's).
The number of users we obtain for each state closely resembles their population (Spearman R $0.97$).

\begin{table*}[tbp]
  \centering
  \caption{ \label{table:logit-table-realworld} %
  Odds ratios obtained by logistic regression. Each column corresponds to a model with a specific set of variables. 
  We indicate with three asterisks the statistically significant correlations ($p < 0.001$). Odds ratios and significance are similar with or without the inclusion of the ``same state'' variables.}
  \footnotesize
  \rowcolors{2}{gray!20}{white}
  \begin{tabular}{llllllllll}

\toprule
Variable name &   \\
\midrule
    
Clinton sup.                  &  0.909*** &  0.884*** &  0.883*** &  0.883*** &  0.883*** &  0.884*** &  0.883*** &  0.883*** &  0.884*** \\
Cross-group                   &  1.172*** &  1.156*** &  1.133*** &  1.165*** &  1.117*** &  1.162*** &  1.163*** &  1.122*** &  1.144*** \\
Clinton sup., Cross-group     &  1.104*** &  1.162*** &  1.163*** &  1.163*** &  1.165*** &  1.164*** &  1.165*** &  1.162*** &  1.164*** \\
Avg. score (author)           &  1.176*** &  1.174*** &  1.175*** &  1.175*** &  1.175*** &  1.174*** &  1.174*** &  1.174*** &  1.173*** \\
Avg. score (target)           &  1.169*** &  1.151*** &  1.151*** &  1.151*** &  1.151*** &  1.151*** &  1.151*** &  1.151*** &  1.151*** \\
Diff. avg. score              &  1.232*** &  1.208*** &  1.209*** &  1.209*** &  1.210*** &  1.209*** &  1.209*** &  1.209*** &  1.209*** \\
Frac. positive (author)       &  1.244*** &  1.208*** &  1.211*** &  1.211*** &  1.207*** &  1.209*** &  1.207*** &  1.212*** &  1.208*** \\
Frac. positive (target)       &  1.192*** &  1.179*** &  1.179*** &  1.179*** &  1.178*** &  1.179*** &  1.178*** &  1.180*** &  1.178*** \\
Diff. frac. positive          &  0.501*** &  0.488*** &  0.488*** &  0.488*** &  0.488*** &  0.488*** &  0.488*** &  0.488*** &  0.488*** \\
Same state                    &  1.245*** &  1.243*** &  1.241*** &  1.241*** &  1.242*** &  1.239*** &  1.240*** &  1.241*** &  1.242*** \\
Same state, Cross-group       &  0.811*** &  0.811*** &  0.810*** &  0.809*** &  0.812*** &  0.815*** &  0.817*** &  0.809*** &  0.814*** \\
Swing state                   &           &    1.018* &           &           &           &           &           &           &           \\
Swing state, Cross-group      &           &    0.965* &           &           &           &           &           &           &           \\
Clinton share                 &           &           &     1.005 &           &           &           &           &           &           \\
Clinton share, Cross-group    &           &           &     1.026 &           &           &           &           &           &           \\
Trump share                   &           &           &           &     0.997 &           &           &           &           &           \\
Trump share, Cross-group      &           &           &           &    0.971* &           &           &           &           &           \\
Non-vote share                &           &           &           &           &    0.977* &           &           &           &           \\
Non-vote share, Cross-group   &           &           &           &           &  1.057*** &           &           &           &           \\
Unemployment                  &           &           &           &           &           &    1.020* &           &           &           \\
Unemployment, Cross-group     &           &           &           &           &           &     0.978 &           &           &           \\
Gini Coefficient              &           &           &           &           &           &           &     1.006 &           &           \\
Gini Coefficient, Cross-group &           &           &           &           &           &           &     0.977 &           &           \\
Median Income                 &           &           &           &           &           &           &           &     1.006 &           \\
Median Income, Cross-group    &           &           &           &           &           &           &           &    1.046* &           \\
High school                   &           &           &           &           &           &           &           &           &     1.002 \\
High school, Cross-group      &           &           &           &           &           &           &           &           &     1.008 \\
\bottomrule
\multicolumn{10}{c}{Odds ratios. * $p < 0.05$, *** $p < 0.001$}
\end{tabular}

\end{table*}

First, we ask ourselves if there could be a homophilic behavior due to geographic proximity, explained by common interests (e.g., local issues), culture, and norms.
To do so, we include in our regression model a dummy variable \emph{``same state''} that indicates whether $u$ and $v$ come from the same US state.
Then, we select a set of macroscopic attributes of each state, that we hypothesize might be related to their online behavior in a political community.
These variables present a basic sketch of the environment of the authors, as represented by state they live in.
In particular we focus on the following political, economic, and demographic variables:
\begin{itemize}
  \item \emph{Swing state}: a dummy variable which indicates whether the author lives in a US state that obtained a 2016 presidential election margin of less than $4\%$ for any candidate.
  \item \emph{Clinton/Trump share}: the shares of votes obtained by these two candidates in the 2016 elections in the state where the author lives.
  \item \emph{Non-vote share}: fraction of the population that did not vote for either of the two major candidates in that state.
  \item \emph{Unemployment}: unemployment rate in 2016 in the state (source: US Bureau of Labor Statistics).
  \item \emph{Gini coefficient}: income inequality in the state, as measured by the Gini coefficient in 2010 (data from the American Community Survey, conducted by the US Census Bureau).
  \item \emph{Median income}: median household income in 2016 (source: American Community Survey).
  \item \emph{High school}: fraction of the population with a high school degree or higher (source: 2013-2017 American Community Survey).
\end{itemize}

We normalize all numerical variables according to quantile normalization, so that they display the same distribution.
The data related to voting behavior refers to the election of November 2016, while our comments are in general gathered from the whole electoral year.
This process is coherent with our hypothesis: is there any difference in behavior in the general population of a US state that could manifest itself also on social media, and that affected the electoral process?

We build a logistic regression model for each one of these variables separately.
In each model, beside the studied variable, we also include the interaction feature between the selected environmental variable and the \emph{cross-group} feature.
This way, we capture whether the selected environmental variable has an effect on the likelihood of a user interacting with another user who supports a different presidential candidate.
We also include in each model all the Reddit-related variables analyzed so far, since they all emerged as significant.
We repeat this analysis for each variable, with and without including the \emph{same state} feature in each of the other models, which may act as a large confounder for the other environmental variables.
We report only results including this variable, but the two cases are quantitatively similar.

Table~\ref{table:logit-table-realworld} shows the odds ratio and the statistical significance obtained by these models.
We consider significant for our hypothesis only models where both the analyzed variable and its interaction with the \emph{cross-group} variable is significant.
Note that the inclusion of environmental factors does not alter significantly the odds ratio and the significance obtained by the community variables, thus further testifying for the robustness of our main results on heterophily and asymmetry.

We summarize the findings obtained via the models in Table~\ref{table:logit-table-realworld} as follows.
\begin{enumerate}[label=(\roman*)]
  \item There is in fact a significant ($p < 0.001$) homophily among users living in the same US state.
  This result suggests that geographical proximity affects the likelihood of political interactions on Reddit.
  Similar results are known for other communities (for instance, the Brexit leave campaign~\cite{bastos2018geographic}).
  A possible explanation is that geographical location subsumes other characteristics---for instance, close-by users could be more likely to share similar interests, and therefore to comment on the same, locally-relevant topic.
  Furthermore, nation-wide political campaigns also involve local matters and candidates: discussing those issues might gather local users.
  \item We find a significant correlation with \emph{non vote}. In particular, states where individuals are most likely to abstain from voting Trump or Clinton in the presidential elections are also those where cross-party interactions on Reddit are most likely ($p < 0.001$). Moreover, in those states, users seem less likely leave a comment, although with less significance ($p < 0.05$).
  This finding is consistent with the idea, well discussed in literature, that exposure to cross-cutting political views is associated with diminished political participation~\cite{bakshy2015exposure,mutz2002consequences,huckfeldt2004disagreement}.%
  \item We find swing states to be less likely to show cross-party interactions, but to foster more homophilic interactions (\mbox{$p < 0.05$}). This result is somewhat surprising, but is consistent with previous findings; e.g., that face-to-face interactions within families decrease when there is political disagreement, which is exacerbated by massive political ads campaigns in swing states~\cite{chen2018effect}.
  \item Higher rates of unemployment seem to lead to the same effect: users coming from states with higher unemployment are more likely to leave a comment. %
  \item We observe a slight variation ($p < 0.05$) between Trump-leaning and Clinton-leaning states: cross-group interactions are less likely in states with higher shares of Trump votes.
  \item The other variables we test do not show a statistically significant correlation: the likelihood of interactions does not seem to be correlated with income inequality or education level.%
\end{enumerate}

\section{Discussion}
In this work, we analyzed cross-group interactions between 
supporters of Trump and those of Clinton on Reddit during the 2016 US presidential elections.
To this aim, we reconstruct the interaction network among these users on the main political discussion community, \R{politics}.
We find that, despite the political polarization,
these groups tend to interact {more} across than among themselves, that is, the network exhibits \emph{heterophily} rather than homophily.
This finding emerges by comparison with a null model of random social interactions, implemented both as a network rewiring that preserves the activity of users,
and as a logistic regression model for link prediction which takes into account possible confounding factors. 

Overall, our findings show that Reddit has been a tool for political discussion between opposing points of view during the 2016 elections.
This behavior is in stark contrast with the echo chambers observed in other polarized debates regarding different topics, on several social media platforms.
While it has been argued that polarization on social media can result in the presence of echo chambers, in which users do not hear opposing views, here we observe the reversed phenomenon: polarization is associated to increased interactions between groups holding opposite opinions. 
However, this relation between polarization and heterophily might not go beyond the digital realm. 
Reportedly, people perceive to encounter more disagreement in online than in offline interactions~\cite{vaccari}.
Further research should be dedicated to understanding whether the heterophily found in this social network is specific about the 2016 presidential elections, or it applies to politics in general, and thus it might be a general feature of the Reddit platform~\citep{cinelli2020echo}. 

\rev{
Several works in the literature have tried quantifying the presence of echo chambers in different social media, although most of them have not studied Reddit.
\citet{conover2011political} analyze 250,000 tweets from the 2010 U.S. congressional midterm elections. They measure the ratio between the observed and expected numbers of links in a random model: they find that users are more likely to interact people with whom they agree, with an odds ratio of $1.2-1.3$ for mentions and $1.7-2.3$ for retweets, thus concluding that the retweet network is highly polarized. In our data, the same measure gives opposite results.
\citet{garimella2018quantifying} quantify the presence of echo chambers in controversial political discussion on Twitter.
Similarly to our methodology, they identify two ingredients that create an echo chamber: the opinion that is shared, and the social network that allows the opinion to echo (operationalized as the follow network).
By looking at the correlation of opinions expressed by a user and opinions exposed to, they confirm the presence of echo chambers (i.e., users with different opinions tend not to follow each other), but only when the topic is controversial.
\citet{bakshy2015exposure} look at the interaction between users and news on Facebook.
Again, the same two main ingredients are the focus of the study: the user opinion (operationalized as self-reported ideological alignment on a liberal-conservative axis), and the social network (friendship network).
Their findings show the presence of homophily in the social network, which is also the main cause of decrease of exposure to ideologically cross-cutting content (compared to the effect of the news feed algorithm).
Given these results, it is possible that the organization of social media as a social network (e.g., Twitter and Facebook), rather than a social forum such as Reddit, fosters the creation of echo chambers.
}
\rev{As previously discussed, there is no unanimous consensus on the effects of echo-chambers in public discourse,
or even on their very existence. 
Ref. \cite{dubois2018echo}, for instance, challenged the hypothesis that echo-chambers actually reduce the content diversity to which social media users are exposed. 
The authors found that many users check news sources different from their usual ones, often offline. They thus argue against generalizing single-media studies to describe the complexity of a high-choice multiple
media environment.
Along the same lines, Ref. \cite{barbera2015tweeting} highlights the importance of taking into account the temporal dimension in the formation (or disgregation) of echo-chambers. 
The authors of Ref. \cite{barbera2015tweeting} found that the exchange of information with respect to 
some controversial topics on Twitter starts and persists as a national conversation participated by different political sides, before sliding into an echo-chamber picture. }

As with any empirical work, our work presents some limitations.
First, Reddit users are not representative of the US population, and instead have strong socio-demographic biases.
Reddit users are more likely to be young males~\citep{duggan2013online}, this finding is also confirmed by informal surveys proposed to users~\citep{agegenderreddit}.
Furthermore, Reddit is much more popular in urban rather than rural areas~\citep{duggan2013online}. 
More importantly, it has been shown that the political leaning and general interests of Reddit users may differ from those of the general population~\citep{redditselfcommunity}.
Even so, the present study is focused on the Reddit community, and does not claim any socio-demographic implication on the general population, besides the ones specifically addressed by the geolocation of users which actually have the opposite causal direction. 
In this respect, a second limitation is evident: the state-level aggregation is coarse-grained, and does not take into consideration differences between areas within the same state (e.g., urban vs rural).
Nevertheless, the state-level is the finest spatial granularity we can reliably infer for a large-enough sample of Reddit users.

Despite these limitations, we find several interesting patterns regarding sociodemographic and environmental factors \rev{associated to an increase in likelihood} of interactions between like-minded individuals~\cite{gentzkow2011ideological}.
To test this hypothesis, we analyzed the effect of different environmental factors by inferring the state of each user according to the information gathered by~\citet{balsamo2019firsthand}.
We observed an effect of \emph{geographical homophily} on the \R{politics} network: interactions between users located in the same state are significantly more likely than random chance.
At the same time, users from the same state are {less} likely to interact when they support different candidates.
Therefore, \rev{we speculate that} while different political views foster interactions in the general case, geographical location \rev{might act} more as a barrier.

Among other environmental factors, we also observed a correlation between the likelihood of cross-group connections and the fraction of the population that abstained from voting.
This finding suggests prudence when defining diversity of exposure as a normative goal.
Similar results were measured through surveys,
with~\citet{mutz2002consequences} arguing that conflicts within one's own social environment can produce ambivalence, which can in turn reduce the intensity of support for one's side.
Interestingly, \citet{mutz2002consequences} observes such decrease even in the absence of new information.
Further empirical evidence is needed to understand this phenomenon and which additional factors may drive it.
We leave this question as important future work.
While weighting our understanding of social media as a dissonating chamber or as an echoic one, we cannot escape the question if dissonance damages the pursuit of common goals for political groups, or if it produces more realistic and less enthusiastic views of the available candidates.

\section*{Acknowledgements}
GDFM and MS acknowledge the support from Intesa Sanpaolo Innovation Center. 
The funder had no role in study design, data collection and analysis, decision to publish, or preparation of the manuscript.

\section*{Author contributions}
GDFM, CM, MS designed the study. GDFM, CM, MS analyzed and discussed the results. GDFM, CM, MS wrote the manuscript. All authors approved the final version of the manuscript.

\section*{Competing interests}
The authors declare no competing interests.

\bibliographystyle{naturemag-doi-custom}
\bibliography{references}
\end{document}